\documentclass[aps,
preprint,
showpacs]{revtex4-1}
\usepackage{amssymb}
\usepackage{amsfonts}
\usepackage{amsmath}
\usepackage{mathrsfs}
\usepackage{graphicx}
\usepackage{flushend}

\begin{document}

\title{Gravitational Coupling from Active Gravity}

\author{Tao Lei}
\author{Zi-Wei Chen}
\author{Zhen-Lai Wang}
\author{Xiang-Song Chen}
\email{cxs@hust.edu.cn}

\affiliation{School of Physics and
MOE Key Laboratory of Fundamental Quantities Measurement,
Huazhong University of Science and Technology, Wuhan 430074, China}

\date{\today}

\begin{abstract}
We attempt to construct a gravitational coupling by pre-selecting an energy-momentum tensor as the source for gravitational field. 
The  energy-momentum tensor we take is a recently derived new expression motivated by joint localization of energy and momentum 
in quantum measurement. This energy-momentum tensor differs from the traditional canonical and symmetric ones, and
the theory we obtain is of an Einstein-Cartan type, but derived from a minimal coupling of a Lagrangian with second-derivative,
and leads to additional interaction between torsion and matter, including the scalar field. 
For the scalar field, the theory can also be derived in the Riemann space-time by a non-minimal coupling. 
Our study gives hint on more general tests of general relativistic effects. 
\pacs{04.20.Cv, 04.20.Fy, 04.50.Kd, 11.30.Cp}
\end{abstract}
\maketitle

\section{Passive and active gravity}
Einstein's general relativity is derived from the equivalence principle, i.e., the local equivalence of inertial and 
gravitational effects on the motion of matter. In this sense, it is a gravitational theory guided by passive gravity. 
The active gravitational source, which can be called the (gravitational)
energy-momentum tensor of matter, is derived {\it post priori} from the 
action constructed under the equivalence principle:
\begin{equation}\label{Tg}
T_g^{\mu\nu} \equiv \frac{2}{\sqrt{g}} \frac{\delta I_M(\varphi,\nabla \varphi)}{\delta g_{\mu\nu}},
\end{equation}
where $\nabla$ is the coordinate-covariant derivative, $g\equiv -det.(g_{\mu\nu})$, and the metric $g_{\mu\nu}$ has 
signature $($$-$$+$$+$$+$$)$. 
It should be noted that the explicit expression of this (active) gravitational energy-momentum tensor 
is not known {\em before} actually computing the variation in Eq. (\ref{Tg}). 
Hence, it is somehow a coincidence that $T_g^{\mu\nu}$ is found to be consistent with the expression obtained by 
symmetrizing the canonical N\"other current in flat space-time with the Belinfante method: 
\begin{equation}\label{Ts}
T_{\rm symm}^{\mu\nu} \equiv T_{\rm cano}^{\mu\nu}+
\frac{i}{2}\partial_\rho \big( \frac{\partial\mathscr{L}}{\partial(\partial_\rho\varphi_a)}\Sigma^{\mu\nu}_{ab}\varphi_b
+\frac{\partial\mathscr{L}}{\partial(\partial_\mu\varphi_a)}\Sigma^{\nu\rho}_{ab}\varphi_b
-\frac{\partial\mathscr{L}}{\partial(\partial_\nu\varphi_a)}\Sigma^{\rho\mu}_{ab}\varphi_b \big).
\end{equation}
Here, $\mathscr{L}=\mathscr{L}(\varphi, \partial \varphi)$ is the standard expression of matter Lagrangian, 
and $\Sigma^{\mu\nu}$ are the spin matrices satisfying the algebra of the homogeneous Lorentz group:
$[\Sigma^{\mu\nu},\Sigma^{\rho\sigma}]=2i(\eta^{\mu[\sigma}\Sigma^{\rho]\nu}-\eta^{\nu[\rho|}\Sigma^{\mu|\sigma]})$,
where $\eta^{\mu\nu}$ is the Minkowski metric tensor. In this paper
$[~]$/$(~)$ means antisymmetrization/symmetrization, and indices inside $|~|$ are excluded from 
symmetrization or antisymmetrization. The canonical energy-momentum tensor is
\begin{equation}\label{Tc}
T_{\rm cano}^{\mu\nu}=-\frac{\partial\mathscr{L}(\varphi,\partial_\mu\varphi)}{\partial(\partial_\mu\varphi)}
{\partial}^\nu\varphi
+\eta^{\mu\nu}\mathscr{L}.
\end{equation}
It is again a somewhat coincidence, that this canonical energy-momentum tensor agrees with the source for the
asymmetric Einstein tensor in the Einstein-Cartan theory \cite{Hehl76}, in which the Belinfante symmetrization
procedure can be reversed to reduce the symmetric metric energy-momentum tensor back to the canonical 
expression.

We may ask a tricky euqation: 
Could it be possible to pre-select an energy-momentum tensor, and require it to be the source for gravity?
Considering $T_{\rm symm}^{\mu\nu}$ or $T_{\rm cano}^{\mu\nu}$, 
for example, there is no criteria to judge which one deserves better to serve as the gravitational source,
thus the competition between the Einstein and Einstein-Cartan theory has to wait for the extremely difficult experimental test. 
The critical problem is that  the conservation laws determine only the conserved charges, not the conserved currents, 
therefore, at the density level, $T_{\rm symm}^{\mu\nu}$ or $T_{\rm cano}^{\mu\nu}$
does not have absolute and independent physical meaning other than acting as the source of gravitational field. 
To construct a gravitational coupling from active gravity, one has to establish a first principle for 
seeking a proper energy-momentum tensor. 

\section{Pre-selected energy-momentum tensor}
In a very recent paper \cite{Wang17}, we proposed a scenario that by examining the local accumulation of energy $E$
and momentum $\vec P$ in a quantum measurement, one can at least define the ``effective" fluxes of energy and momentum:
\begin{equation}\label{Teff}
T^{i0}_{\rm eff} \cdot dS_i \equiv \frac {d E}{dt}, ~ ~ T^{ij}_{\rm eff} \cdot dS_i \equiv \frac {d P^j}{dt}.
\end{equation}
Virtually, for any locally measurable conserved charge $Q$, one can define an effective current $\vec j_{\rm eff} $ (or more
exactly the projected component $j_{n{\rm ,eff}}$):
\begin{equation}
j_{n{\rm ,eff}} \cdot dS= \vec j_{\rm eff}  \cdot d \vec S \equiv \frac{dQ}{dt}.
\end{equation}

Strictly speaking, it is not clear and not for sure why such ``effective" current should have fundamental physical significance.
But as explained in Ref. \cite{Wang17}, by requiring joint localization of all conserved quantities in a quantum measurement, 
the corresponding effective currents can be correlated, which thus sets a practical constraint on the allowed expressions of  
such currents. Strikingly and interestingly, neither  $T_{\rm symm}^{\mu\nu}$ nor $T_{\rm cano}^{\mu\nu}$ can satisfy such 
a correlation, thus they are not the correct description of local fluxes of energy and momentum  in quantum measurements.
Certainly, this does not tell that $T_{\rm symm}^{\mu\nu}$ or $T_{\rm cano}^{\mu\nu}$
cannot be proper candidate for gravitational source, but it is interesting and worthwhile to explore 
what expression $T_{\rm eff}^{\mu\nu}$ can take and what theory we get by setting $T_{\rm eff}^{\mu\nu}$ as the gravitational source. 

For a free field $\varphi$ \cite{note}, describing either an elementary particle or a composite particle, 
an elegant expression of $T_{\rm eff}^{\mu\nu}$ can be found \cite{Wang17}:
\begin{equation}\label{Tnew}
T^{\mu\nu}_{\rm eff}=-\frac{\partial\mathscr{L}}{\partial(\partial_\mu\varphi)}\overleftrightarrow{\partial}^\nu\varphi
=-\frac{1}{2} \big( \frac{\partial\mathscr{L}}{\partial(\partial_\mu\varphi)}{\partial}^\nu\varphi
-\partial^\nu\frac{\partial\mathscr{L}}{\partial(\partial_\mu\varphi)}\varphi \big),
\end{equation}
where $\overleftrightarrow{\partial}^\nu=\frac{1}{2}(\overrightarrow{\partial}^\nu-\overleftarrow{\partial}^\nu)$.
If one considers further the measurement of angular momentum, a proper expression satisfying the correct correlation 
of currents can also be derived \cite{Wang17}:
\begin{eqnarray}\label{Mnew}
M^{\mu\nu\rho}_{\rm eff}&=&\big(x^\mu T_{\rm eff}^{\rho\nu}-x^\nu T_{\rm eff}^{\rho\mu}\big)
+\big(-i\frac{\partial\mathscr{L}}{\partial(\partial_\rho\varphi_a)} \Sigma ^{\mu\nu}_{ab}\varphi_b 
-\frac{1}{2}\eta^{\rho\nu}\frac{\partial\mathscr{L}}{\partial(\partial_\mu\varphi_a)}\varphi_a
+\frac{1}{2}\eta^{\rho\mu}\frac{\partial\mathscr{L}}{\partial(\partial_\nu\varphi_a)}\varphi_a \big)
\nonumber \\
&\equiv& L^{\mu\nu\rho}_{\rm eff}+s^{\mu\nu\rho}_{\rm eff}.
\end{eqnarray}
$T^{\mu\nu}_{\rm eff}$ and $M^{\mu\nu}_{\rm eff}$ differ from their conventional counterparts by total-divergence terms, 
which change neither the conservation law nor the conserved charge. Note that in $M^{\mu\nu\rho}_{\rm eff}$, the spin current 
$s^{\mu\nu\rho}_{\rm eff}$ contains an extra term which exists also for the scalar field. 
We will see that this extra spin current can couple to space-time torsion. 

It is a great advantage for studying gravitational coupling that $T^{\mu\nu}_{\rm eff}$ and $M^{\mu\nu}_{\rm eff}$ can be derived as 
N\"other currents via a new Lagrangian density $\mathscr{L}_{\rm new} (\varphi,\partial\varphi,\partial \partial\varphi)$, 
which has a critical property of vanishing upon applying the equation of motion:
 \begin{eqnarray}\label{Lnew}
\mathscr{L}_{\rm new} (\varphi,\partial\varphi,\partial\partial\varphi)=\frac{1}{2}\varphi\big(\frac{\partial\mathscr{L}}{\partial\varphi}
-\partial_\mu\frac{\partial\mathscr{L}}{\partial(\partial_\mu\varphi)}\big).
\end{eqnarray}
In general, it contains second derivative (except for the Dirac field), 
and differs from the standard $\mathscr{L}$ by a surface term:
 \begin{eqnarray}\label{Lnew'}
\mathscr{L}_{\rm new} (\varphi,\partial\varphi,\partial\partial\varphi)=\mathscr{L}(\varphi,\partial\varphi)
-\frac{1}{2}\partial_\mu \big(\varphi\frac{\partial\mathscr{L}}{\partial(\partial_\mu\varphi)}\big).
\end{eqnarray}

Eqs. (\ref{Lnew}) and (\ref{Lnew'}) prepare us to construct a gravitational coupling
with the guidance that a covariant extension of the effective energy-momentum tensor in
Eq. (\ref{Tnew}) should act as the gravitational source. 

\section{Minimal versus non-minimal coupling}

The simplest way to switch on the gravitational interaction is by minimal coupling, namely, replacing the Lorentz-covariant 
quantities in the flat space-time action with their counterparts which are covariant under arbitrary coordinate transformation. 
Let us also make this attempt, starting with the
Lagrangian in Eq. (\ref{Lnew}) which can give the desired N\"other currents. However, nothing new will be gained in the 
Riemann space-time, in which the covariant Gauss' theorem tells us that the surface term in Eq. (\ref{Lnew'}) is irrelevant. 
So let us try instead the Riemann-Cartan space-time with non-zero torsion.

To facilitate the inclusion of spinor field, we use the tetrad formalism, in which the energy-momentum tensor and 
spin tensor can be conveniently evaluated as 
\begin{equation}\label{tetradTS}
\mathbf{T}^{\mu\nu} \equiv  
e^{a\nu} \cdot \big( {\mathbf{T}^\mu}_a \equiv  \frac{1}{e}\frac{\delta I_m}{\delta{e^a}_\mu} \big),
~
\mathbf{s}^{\mu\nu\rho} \equiv 
e^{a\mu}{e_b}^\nu \cdot \big( {\mathbf{s}_a}^{b\rho} \equiv -\frac{2}{e}\frac{\delta I_m}{\delta{\omega^a}_{b\mu}}\big).
\end{equation}
The tetrad quantities are defined by 
$g_{\mu\nu}={e^a}_\mu {e^b}_\nu\eta_{ab}$, $\eta_{ab}={e_a}^\mu {e_b}^\nu g_{\mu\nu}$, $e=det({e^a}_\mu)$.
Latin and Greek letters denote Lorentz and coordinate indices, respectively. 
${\omega^a}_{b\mu}$ is the spin connection as appear in the covariant derivative, e.g., of a Lorentz-vector
$A^a={e^a}_\mu A^\mu$:
$\nabla_\mu A^a=\partial_\mu A^a+{\omega^a}_{b\mu}A^b$.

In the Einstein-Cartan theory, the matter action $I_{m}$ is constructed via minimal coupling of the standard Lagrangian:
\begin{equation}
I_m(\varphi)=\int d^{4}x\mathscr{L}(\varphi,\partial\varphi)
\rightarrow I_m^{\rm EC} (e, \omega, \varphi)=\int d^{4}xeL(\varphi,\nabla\varphi); 
\end{equation}
and in our model, it is constructed via minimal coupling of the Lagrangian in Eq. (\ref{Lnew}):
\begin{equation}\label{Inew}
I_{m}^{\rm new}(\varphi)=\int d^{4}x\mathscr{L}_{\rm new}(\varphi,\partial\varphi,\partial^2\varphi) 
\rightarrow I_{m}^{\rm new} (e, \omega, \varphi)=\int d^{4}x e L_{\rm new} (\varphi,\nabla\varphi,\nabla\nabla \varphi) .
\end{equation}

When varying the total action of matter and geometry $I=I_m +I_g$ with respect to tetrad and spin connection independently, 
one can obtain equations of motion of gravitation:
\begin{equation}
{G^\mu}_a=8\pi{G}{\mathbf{T}^\mu}_a,~
{\mathcal S _a}^{b\mu}=-4\pi{G}{\mathbf{s}_a}^{b\mu}.
\end{equation} 
In Riemann-Cartan space-time, the gravitational action is still taken as $I_g={\int}d^{4}xe\frac{R}{16\pi{G}}$, but 
the Einstein tensor $G^{\mu\nu}=e^{a\nu}{G^\mu}_a$ is generally asymmetric. 
${\mathcal S}^{\mu\nu\rho}(=e^{a\mu}{e_b}^\nu {{\mathcal S}_a}^{b\rho}=S^{\mu\nu\rho}+g^{\mu\rho}S^\nu-g^{\nu\rho}S^\mu)$ 
is called the modified torsion tensor, and the torsion tensor is
$ {S_{\mu\nu}}^\rho=\Gamma_{[\mu\nu]}^\rho={\omega^\rho}_{[\mu\nu]}+{e_a}^\rho\partial_{[\nu}{e^a}_{\mu]}$.
Its trace ${S_{\mu\nu}}^\nu \equiv S_\mu$ is defined as the torsion vector.

We will show in the next section with the specific scalar, spinor, and vector fields that in our model the 
energy-momentum tensor is exactly the covariant extension of $T_{\rm eff}^{\mu\nu}$ in Eq. (\ref{Tnew}). 
Thus, guided by ``active'' gravity, we can indeed arrive at a sensible model of gravitational interaction, 
derived by minimal coupling of a specific and valid Lagrangian in flat space-time.

It should be remarked that our model can also be viewed as one with non-minimal coupling. 
The point is that the equivalence between $\mathscr{L}$ and $\mathscr{L}_{\rm new}$ in flat or even Riemann space-time 
is lost in the  Riemann-Cartan space-time. 
Let us compare the matter action $I^{\rm new}_m(e, \omega, \varphi)$ in our model with that in the Einstein-Cartan theory:
\begin{eqnarray}\label{compare}
I^{\rm new}_m(e, \omega, \varphi)&=&
{\int}d^{4}xeL_{\rm new}(\varphi,\nabla\varphi,\nabla\nabla \varphi)
={\int}d^{4}xe\frac{1}{2}\varphi(\frac{\partial L}{\partial\varphi}
-\nabla_a\frac{\partial{L}}{\partial(\nabla_a\varphi)})\nonumber\\
&=&{\int}d^{4}xe[ {L}-\frac{1}{2}\nabla_a(\varphi \frac{\partial{L}}{\partial(\nabla_a\varphi)})]
=I_m^{\rm EC}-{\int}d^{4}xe \varphi\frac{\partial{L}}{\partial(\nabla_a\varphi)}S_a,
\end{eqnarray}
where the relation $\nabla_a A^a=\widetilde{\nabla}_a A^a+2S_a A^a$ is used, with
$\widetilde{\nabla}_a$ the covariant derivative built with Christoffel connection. 

If one regards ${L}_{\rm new}$ as the fundamental Lagrangian, our model is a minimal-coupling one. 
On the other hand, if priority is assigned to $L$ which contains only the first derivative, then 
as the last expression of Eq. (\ref{compare}) indicates, our model can be viewed as an extension of Einstein-Cartan
theory by including a non-minimal coupling between matter and torsion.
Such ambiguity of minimal coupling in the Einstein-Cartan space-time had already been discussed before 
\cite{Kazm09}. From our analysis in this paper, 
even the original Einstein-Cartan action can be regarded as one of non-minimal coupling,
were  our ${L}_{\rm new}$ given a priority.

\section{Explicit construction for scalar, spinor, and vector fields}
In this section, we present explicitly the active-gravity-guided gravitational coupling for the scalar, spinor, and vector fields,
respectively, and verify that the gravitational source is indeed the covariant extension of the ``effective'' currents as 
required to properly describe the fluxes of conserved quantities in quantum measurement. 

\subsection{Scalar field}\label{scalar}
The proper flat space-time Lagrangian for the free scalar field $\phi$ in the form of Eq. (\ref{Lnew}) is:
 \begin{equation}
{\mathscr L}_{\phi }^{\rm new}=\phi(\partial_\mu\partial^\mu-m^2)\phi.
\end{equation}
In the spirit of Eq. (\ref{Inew}), the corresponding action in Riemannian-Cartan space-time is 
\begin{subequations}
\begin{align}
I_\phi^{\rm new} (e,\omega,\phi)&={\int}d^{4}xe\frac{1}{2}[\phi(\nabla_a\nabla^a-m^2)\phi] 
\equiv {\int}d^{4}xe {L}_{\phi }^{\rm new} (e,\omega,\phi)\\
&={\int}d^{4}xe[-\frac{1}{2}\nabla_a\phi\nabla^a\phi-\frac{1}{2}m^2\phi^2
+\phi(\nabla_a\phi) S^a]  
\equiv {\int}d^{4}xe {L}_{\phi }^{\rm 'new} (e,\omega,\phi).
\end{align}
\end{subequations}
As we noted above, the first expression is of the form of minimal coupling, while the second expression 
displays a non-minimal coupling, which indicates clearly that the scalar field interacts with torsion. This 
action gives the equation of motion for $\phi$:
\begin{equation}\label{phi-our}
\frac{1}{2}\nabla_a\nabla^a\phi+\frac{1}{2}\stackrel{\ast~}{\nabla_a}\stackrel{\ast~}{\nabla^a}\phi-m^2\phi=0,
\end{equation}
where $\stackrel{\ast~}{\nabla^\mu}=\nabla^\mu-2S^\mu$ is called the modified divergence. It compares to 
\begin{equation}\label{phi-EC}
\stackrel{\ast~}{\nabla_a}{\nabla^a}\phi-m^2\phi=0
\end{equation}
in the Einstein-Cartan theory with the action
\begin{equation}
I_\phi (e,\omega,\phi)={\int}d^{4}xe[-\frac{1}{2}\nabla_a\phi\nabla^a\phi-\frac{1}{2}m^2\phi^2] .
\end{equation}

By computing Eq. (\ref{tetradTS}) explicitly and applying the covariant equation of motion, we get 
the tetrad energy-momentum tensor and tetrad spin tensor of the scalar field:
\begin{equation}\label{Tphi}
\mathbf{T}^{\mu\nu}=\frac{1}{2}\stackrel{\ast~}{\nabla^\mu}\phi\nabla^\nu\phi-\frac{1}{2}\phi\nabla^\nu\nabla^\mu\phi
+g^{\mu\nu}L_\phi^{\rm new},
~
\mathbf{s}^{\mu\nu\rho}=\frac{1}{2}(g^{\nu\rho}\phi\nabla^\mu\phi-g^{\mu\rho}\phi\nabla^\nu\phi).
\end{equation}
In the flat space-time limit, they reduce to the forms in Eqs. (\ref{Tnew}) and (\ref{Mnew}):
\begin{equation}\label{phi-flat}
\mathbf{T}_{(0)}^{\mu\nu}=T^{\mu\nu}_{\rm eff}=\partial^\mu\phi\overleftrightarrow{\partial}^\nu\phi,
~
\mathbf{s}_{(0)}^{\mu\nu\rho}={s}^{\mu\nu\rho}_{\rm eff}=\eta^{[\nu|\rho}\phi\partial^{|\mu]}\phi.
\end{equation}

\subsection{Spinor field}
For the Dirac spinor field, our model actually coincides with the Einstein-Cartan theory. The reason is that the usual 
expression of Dirac Lagrangian in flat space-time:
 \begin{equation}
{\mathscr L} _\psi =\frac{i}{2}\overline{\psi}\gamma^\mu\partial_\mu\psi
-\frac{1}{2}m\overline{\psi}\psi+h.c.,
\end{equation}
is itself zero when applying the Dirac equation, thus is already of the form in Eq.(\ref{Lnew}). So, when going to  
Riemannian-Cartan space-time, our approach gives the same action of the Einstein-Cartan type:
\begin{equation}
I_\psi(e,\omega,\psi,\overline{\psi})={\int}d^{4}x e
\big( \frac{i}{2}\overline{\psi}\gamma^a\nabla_a\psi-\frac{1}{2}m\overline{\psi}\psi+h.c.\big )
\equiv {\int}d^{4}x e L_\psi,
\end{equation}
where $\nabla_a\psi={e_a}^\mu(\partial_\mu+\frac{i}{4}\sigma_{ab}{\omega^{ab}}_\mu)\psi.$ 
The equations of motion are
\begin{equation}\label{psi-EC}
\frac{i}{2}\gamma^a\nabla_a\psi+\frac{i}{2}\gamma^a\stackrel{\ast~~}{\nabla_a}\psi-m\psi=0,
~~
\frac{i}{2}\nabla_a\overline{\psi}\gamma^a+\frac{i}{2}\stackrel{\ast~~}{\nabla_a}\overline{\psi}\gamma^a+m\overline{\psi}=0
\end{equation}
which resemble more of Eq.  (\ref{phi-our}) in our model, rather than Eq.  (\ref{phi-EC}) in the Einstein-Cartan theory. 

The tetrad energy-momentum tensor and tetrad spin tensor of the Dirac field are trivial covariant extension 
of the familiar canonical expressions:
\begin{equation}
\mathbf{T}^{\mu\nu}=-\frac{i}{2}\overline{\psi}\gamma^\mu\nabla^\nu\psi+h.c. +g^{\mu\nu} L_\psi, 
~
\mathbf{s}^{\mu\nu\rho}=
\frac{1}{4}\overline{\psi}(\gamma^\rho\sigma^{\mu\nu}+\sigma^{\mu\nu}\gamma^\rho)\psi.
\end{equation}

\subsection{Vector field}
For a massive vector field, the proper flat space-time Lagrangian in the form of Eq. (\ref{Lnew}) is:
 \begin{equation}
{\mathscr L}_A^{\rm new}=\frac 12A_\nu(\partial_\mu F^{\mu\nu}-m^2A^\nu).
\end{equation}
Following again Eq. (\ref{Inew}), we get the corresponding action in Riemann-Cartan space-time:
\begin{subequations}
\begin{align}
I_A^{\rm new}(e,\omega,A)&=\frac{1}{2}{\int}d^{4}x e[A_b(\nabla_a F^{ab}-m^2A^b)]
\equiv {\int}d^{4}x e L_A^{\rm new}(e,\omega,A) 
\\
&={\int}d^{4}x e(-\frac{1}{4}F^{ab}F_{ab}-\frac{1}{2}{m^2}A^2+A^{b}F_{ab}S^{a})
\equiv {\int}d^{4}x e L_A^{\rm 'new}(e,\omega,A) .
\end{align}
\end{subequations}
Viewed either as a minimal-coupling one or non-minimal-coupling one, this action gives a spin-torsion 
interaction of both the traditional type as in Einstein-Cartan theory, and a new type related to the 
extra spin current in Eq. (\ref{Mnew}). It gives the equation of motion:
\begin{equation}\label{A-our}
\frac{1}{2}\nabla_a F^{ab}+\frac{1}{2}\stackrel{\ast~}{\nabla_a}\stackrel{\ast~~}{F^{ab}}-m^2A^b=0,
\end{equation}
compared to 
\begin{equation}\label{A-EC} 
I_A(e,\omega,A)={\int}d^{4}x e(-\frac{1}{4}F^{ab}F_{ab}-\frac{1}{2}{m^2}A^2),~~
\stackrel{\ast~}{\nabla_a}{F^{ab}}-m^2A^b=0
\end{equation}
in the Einstein-Cartan theory.  Again, it is Eq. (\ref{A-our}) of our model rather than Eq. (\ref{A-EC}) of the original
Einstein-Cartan theory that resembles more of Eq. (\ref{psi-EC}). Together with the similarity between Eqs. (\ref{phi-our})
and (\ref{psi-EC}), this might be regarded as a hint that our chosen 
Lagrangian in Eq. (\ref{Lnew}) is better for minimal coupling in the Einstein-Cartan space-time. 

Inserting our new action into Eq. (\ref{tetradTS}) and applying the covariant equation of motion, 
we get the tetrad energy-momentum tensor and tetrad spin tensor of the massive vector field:
\begin{equation}\label{TA}
\mathbf{T}^{\mu\nu}=\frac{1}{2}\stackrel{\ast~~}{F^{\mu\rho}}\nabla^\nu A_\rho-\frac{1}{2}\nabla^\nu F^{\mu\rho}A_\rho
+g^{\mu\nu}L_A^{\rm new}, 
~
\mathbf{s}^{\mu\nu\rho}
=A^{[\mu} F^{\nu]\rho}+A^{[\mu}\stackrel{\ast~~~}{F^{\nu]\rho}}+g^{[\nu|\rho} A_\lambda F^{|\mu]\lambda},
\end{equation}
where $\stackrel{\ast~~}{F^{\mu\nu}}\equiv \stackrel{\ast~}{\nabla^\mu} A^\nu-\stackrel{\ast~}{\nabla^\nu} A^\mu$.

Again, in the flat space-time limit, they reduce to the forms in Eqs. (\ref{Tnew}) and (\ref{Mnew}):
\begin{equation}
\mathbf{T}_{(0)}^{\mu\nu}=T^{\mu\nu}_{\rm eff}=F^{\mu\rho}\overleftrightarrow{\partial}^{\nu}A_\rho,
~
\mathbf{s}_{(0)}^{\mu\nu\rho}={s}^{\mu\nu\rho}_{\rm eff}=2A^{[\mu|}F^{|\nu]\rho}+\eta^{\rho[\nu|}F^{|\mu]\alpha}A_\alpha.
\end{equation}

\section{Metric-torsion formulation}

Since our model differs from the Einstein-Cartan theory for the scalar and vector fields only, it suffices to work 
in the metric-torsion formalism, which offers a cross-check as well as some new light on how our new energy-momentum 
and spin tensors participate in gravity.  In this formalism, the independent geometric variables can be taken as the 
metric and contortion tensor
${K_{\mu\nu}}^\rho={S^\rho}_{\mu\nu}+{S^\rho}_{\nu\mu}+{S_{\nu\mu}}^\rho$,
and variation of the gravitational action $I_g(g,K) ={\int} d^{4}x \sqrt{g} \frac{1}{16\pi{G}}R$
can be put in the form:
\begin{equation}
\delta I_g (g,K) ={\int}d^{4}x\sqrt{g}\frac{1}{16\pi{G}}\{[-G^{\mu\nu}+\stackrel{\ast}{\nabla}_\rho (M^{\mu\nu\rho}
+ M^{\mu\rho\nu}+ M^{\nu\rho\mu})]\delta{g}_{\mu\nu}+2{M_\rho}^{\mu\nu}\delta{K_{\mu\nu}}^\rho\}.
\end{equation}
Then, by writing the variation of the matter action as
\begin{equation}
\delta{I}_m(g,K,\varphi)=\int{d}^4x\sqrt{g}
\frac{1}{2}(\mathbf{T}_g^{\mu\nu}\delta{g}_{\mu\nu}+{\mathbf{s}_\rho}^{\mu\nu}\delta{K_{\mu\nu}}^\rho),
\end{equation}
one can get the gravitational equations of motion: 
\begin{equation} \label{GS}
{G}^{\mu\nu}-\stackrel{\ast}{\nabla}_\rho ({\mathcal S}^{\mu\nu\rho}
+ {\mathcal S}^{\mu\rho\nu}+ {\mathcal S}^{\nu\rho\mu})=8\pi G\mathbf{T}_g^{\mu\nu},
~
{{\mathcal S}_\rho}^{\mu\nu}=-4\pi{G}{\mathbf{s}_\rho}^{\mu\nu}.
\end{equation}
Here $\mathbf{T}_g^{\mu\nu}$ is the metric energy-momentum tensor in Riemann-Cartan space-time, 
which is still symmetric. Noting that the equation for torsion is an algebraic one, thus Eq. (\ref{GS}) can be deformed as 
\begin{equation}
{G}^{\mu\nu}=8\pi{G}\mathbf{T}^{\mu\nu},
\end{equation}
and we can identify the source of the Einstein tensor as the energy-momentum tensor:
\begin{equation}\label{rB}
\mathbf{T}^{\mu\nu}=\mathbf{T}_g^{\mu\nu}-\frac{1}{2}\stackrel{\ast}{\nabla}_\rho(\mathbf{s}^{\mu\nu\rho}
+\mathbf{s}^{\mu\rho\nu}+ \mathbf{s}^{\nu\rho\mu}).
\end{equation}
Comparing with Eq. (\ref{Ts}), this can be recognized as a kind of ``reversed'' Belinfante construction, 
thus $\mathbf{T}^{\mu\nu}$ is expected to be a canonical-type energy-momentum tensor, which is generally asymmetric
for a field with spin. 

Specifically, for the scalar field, the action is now written as
\begin{subequations}
\begin{align}
I_\phi(g,K,\phi)&={\int}d^{4}x\sqrt{g}[\frac{1}{2}\phi(\nabla_\mu\nabla^\mu-m^2)\phi]
\equiv {\int}d^{4}x\sqrt{g} L_\phi ^{\rm new}(g,K,\phi) \\
&={\int}d^{4}x\sqrt{g}(-\frac{1}{2}\nabla_\mu\phi\nabla^\mu\phi-\frac{1}{2}m^2\phi^2+\phi\nabla^\mu\phi S_\mu)
\equiv  {\int}d^{4}x\sqrt{g} L_\phi ^{\rm 'new}(g,K,\phi),
\end{align}
\end{subequations}
with the metric energy-momentum tensor and (contortion) spin tensor computed to be
\begin{equation}
{\mathbf{T}_g}^{\mu\nu}=\stackrel{\ast~~}{\nabla^{(\mu}}\phi\nabla^{\nu)}\phi+g^{\mu\nu}L_\phi ^{\rm 'new},
~
\mathbf{s}^{\mu\nu\rho}=g^{[\nu|\rho}\phi\nabla^{|\mu]}\phi.
\end{equation}
Note that in the usual Einstein-Cartan theory, the scalar field has no spin current and does not couple to torsion. 
But in our model, the scalar field acquires a non-zero spin current, which by Eq. (\ref{rB}) converts the metric 
energy-momentum tensor to be our desired one:
\begin{equation}
\mathbf{T}^{\mu\nu}=\mathbf{T}_g^{\mu\nu}-\frac{1}{2}\stackrel{\ast}{\nabla}_\rho(\mathbf{s}^{\mu\nu\rho}
+\mathbf{s}^{\mu\rho\nu}+ \mathbf{s}^{\nu\rho\mu})
=\frac{1}{2}\stackrel{\ast~}{\nabla^\mu}\phi\nabla^\nu\phi-\frac{1}{2}\phi\nabla^\nu\nabla^\mu\phi 
+g^{\mu\nu}L_\phi ^{\rm new}.
\end{equation}
It differs from that in the Einstein's theory, and agrees with Eq. (\ref{Tphi}) in the tetrad formalism. 

For the vector field, the action is now written as
\begin{subequations}
\begin{align}
I_A^{\rm new}(g,K,A)&={\int}d^{4}x \sqrt{g} \frac{1}{2}[A_\nu(\nabla_\mu F^{\mu\nu}-m^2A^\nu)]
\equiv {\int}d^{4}x \sqrt{g} L_A^{\rm new}(g,K,A) \\
&={\int}d^{4}x \sqrt{g}(-\frac{1}{4}F^{\mu\nu}F_{\mu\nu}-\frac{1}{2}{m^2}A^2+A^{\nu}F_{\mu\nu}S^{\mu})
\equiv {\int}d^{4}x \sqrt{g} L_A^{\rm 'new}(g,K,A) ,
\end{align}
\end{subequations}
and the metric energy-momentum tensor and (contortion) spin tensor are computed to be
\begin{equation}
{\mathbf{T}_g}^{\mu\nu}=\stackrel{\ast~~~}{F^{(\mu|\rho}}F^{|\nu)}_{~~\rho}+m^2A^\mu A^\nu+g^{\mu\nu}L_A^{\rm 'new},
~~
\mathbf{s}^{\mu\nu\rho}
=A^{[\mu} F^{\nu]\rho}+A^{[\mu}\stackrel{\ast~~~}{F^{\nu]\rho}}+g^{[\nu|\rho} A_\lambda F^{|\mu]\lambda}
\end{equation}
Because the vector field acquires both the traditional spin current and our ``new'' spin current, the explicit computation 
of  Eq. (\ref{rB}) is a little lengthy, but straightforward, which gives exactly 
the same expression as Eq. (\ref{TA}) in the tetrad formalism:
\begin{equation}
\mathbf{T}^{\mu\nu}=\mathbf{T}_g^{\mu\nu}-\frac{1}{2}\stackrel{\ast}{\nabla}_\rho(\mathbf{s}^{\mu\nu\rho}
+\mathbf{s}^{\mu\rho\nu}+ \mathbf{s}^{\nu\rho\mu})
=\frac{1}{2}\stackrel{\ast~~}{F^{\mu\rho}}\nabla^\nu A_\rho-\frac{1}{2}\nabla^\nu F^{\mu\rho}A_\rho
+g^{\mu\nu}L_A^{\rm new} .
\end{equation}

\section{Alternative theory for the scalar field}

It is notable that although the scalar field in our model acquires a spin current, its ``effective'' 
energy-momentum tensor remains symmetric. This suggests that we may conjecture a model
within the Riemann space-time to assign our effective energy-momentum tensor as the gravitational source. 
This goal, however, can only be possibly achieved with a substantially non-minimal coupling, which cannot be 
converted into a total divergence and thus can survive in the Riemann space-time. 
We find that the following action makes such a model:
\begin{equation}
I(g,\phi)=I_g(g)+I_{\phi}(g,\phi)={\int}d^{4}x\sqrt{g} \frac{\widetilde{R}}{16\pi G}
+{\int}d^{4}x\sqrt{g}(-\frac{1}{2}\widetilde{\nabla}_\mu\phi\widetilde{\nabla}^\mu\phi
-\frac{1}{2}m^2\phi^2+\frac{1}{8}\widetilde{R}\phi^2).
\end{equation}
Here $\widetilde{X}$ denotes a quantity $X$ in the Riemann space-time. 
For example, $\widetilde{\Gamma}_{\mu\nu}^\rho$ signifies the Christoffel connection. 
The action contains a non-minimal $R\phi^2$ interaction, with the coupling coefficient fixed to be 1/8. 
It gives an energy-momentum tensor of the scalar field
 \begin{equation}
 \mathbf{T}^{\mu\nu}=\frac{2}{\sqrt{g}}\frac{\delta{I_{\phi}(g,\phi)}}{\delta{g_{\mu\nu}}}
 =\frac{1}{2}(\widetilde{\nabla}^\mu\phi\widetilde{\nabla}^\nu\phi-\phi\widetilde{\nabla}^\mu\widetilde{\nabla}^\nu\phi)
-\frac{1}{4}R^{\mu\nu}\phi^2.
\end{equation}
In the limit of flat space-time, this agrees with Eq. (\ref{phi-flat}) of Section \ref{scalar}.

Note that the scalar field here is not necessarily a fundamental field as employed in cosmological models. 
It can as well describe a spin-less composite particle. 

\section{Discussion}
As the Einstein-Cartan theory does, our model leads to spin contact interaction, but of more extensive structures. 
Certainly,  the test of such contact interaction has to await extremely precise measurements \cite{Duan16},
but it does not mean that our discussion is purely academic. 
In fact, the most valuable light which our study might shed on the test of gravitational effect, 
independent of the possible merit of our gravitational-coupling model itself, 
is that if $T_g^{\mu\nu}$ differs from $T_{\rm eff}^{\mu\nu}$,  then some peculiar effect may occur.
For example, in Einstein's general relativity, it is $T_g^{\mu\nu}$ that couples to
gravity, while during a quantum measurement 
the effective fluxes of energy and momentum of a quantum wave is dictated by  
$T_{\rm eff}^{\mu\nu}$ in Eq. (\ref{Tnew}), which is indeed different from $T_g^{\mu\nu}$.
This may lead to some kind of non-local violation of universality of free fall for quantum waves, 
and one may conjecture a possible gravitational discrimination of freely falling atomic waves of different species.     

In this paper, we have worked with massive vector field to avoid the discussion of gauge invariance, which is highly tricky 
and controversial \cite{Lead14}.  In the Riemann-Cartan space-time, the minimal coupling between gauge field  
and torsion is gauge-dependent and hence often abandoned \cite{Hojm78}. 
Nethertheless, the recent technique to construct gauge-invariant gluon spin 
\cite{Lead14, Chen08, Chen09}
may be adopted to build a gauge-invariant minimal coupling of photon or gluon to torsion. 
Thus, exploring the interaction between torsion and gauge particles is of vital importance and interest for 
the fundamental aspects of not only gravity, but also gauge theory. 

We thank Wei Xu and De-Tian Yang for helpful discussion. 
This work is supported by the China NSF via Grants No. 11535005 and No. 11275077.


\begin{thebibliography}{99}
\bibitem{note} As noted in \cite{Wang17}, the inclusion of interaction between matter fields is rather tricky, 
and is omitted in the study of matter-gravity coupling in this paper.
\end{thebibliography}
\end{document}